\documentclass[10pt,journal,letterpaper]{IEEEtran}
\IEEEoverridecommandlockouts
\usepackage[utf8]{inputenc}
\usepackage[T1]{fontenc}
\usepackage{newunicodechar}
\newunicodechar{∼}{\textasciitilde{}}
\usepackage{orcidlink}
\usepackage{setspace}
\usepackage{booktabs}
\usepackage{tabularx}
\usepackage{siunitx}
\usepackage{subcaption}
\IEEEoverridecommandlockouts
\usepackage{fancyhdr}
\usepackage[numbers, sort&compress]{natbib}
\usepackage{url}
\usepackage{graphicx}
\usepackage{float}
\usepackage{enumerate}
\usepackage{hyperref}
\usepackage{amsmath,amssymb,amsfonts}
\usepackage{algorithmic}
\usepackage[ruled,vlined]{algorithm2e}
\usepackage{graphicx}
\usepackage{array}     
\usepackage{amsmath}   
\usepackage{textcomp}
\usepackage{xcolor}
\usepackage{physics}

\usepackage[font=footnotesize, skip=0pt,compatibility=false]{caption}
\usepackage{subcaption}
\usepackage{comment}
\usepackage{tikz}
\usetikzlibrary{positioning, shapes, arrows.meta}
\usepackage{booktabs}
\usepackage{multirow}
\def\BibTeX{{\rm B\kern-.05em{\sc i\kern-.025em b}\kern-.08em
    T\kern-.1667em\lower.7ex\hbox{E}\kern-.125emX}}
\begin{document} 
\bstctlcite{MyBSTcontrol}
\title{Quantum Secure Time Transfer for Satellites}
\author{Ravi Singh Adhikari$^{*}$, Anju Rani$^{*}$, Aman Gupta$^{*}$, Xiaoyu~Ai$^{*}$, 
and~Robert~Malaney$^*$ 
\thanks{$^*$Ravi Singh Adhikari, Anju Rani, Aman Gupta, Xiaoyu Ai, and Robert Malaney are with the School of Electrical Engineering and Telecommunications, University of New South Wales, Sydney, Australia.}
}

\maketitle
\thispagestyle{empty}
\pagestyle{empty}
\begin{abstract}
We experimentally demonstrate an entanglement-based Quantum-Secure Time
Transfer  (QSTT) system in an emulated low Earth orbit satellite-to-ground channel using a \mbox{type-0} Sagnac-based entangled-photon source.
Our new QSTT system delivers a finite Quantum Key Distribution (QKD) key rate of approximately $3$~bits~$\text{s}^{-1}$ with a $10^{-10}$ security parameter, providing a $30$-fold increase in the QKD key rate compared to the state-of-the-art entanglement-based QKD system delivered by the Micius satellite. Beyond this high key rate outcome,
novel to our QSTT system is a GPS-free clock synchronization, optimized use of QKD, embedded post-quantum security, and obfuscation of the  system configuration via use of a pre-shared key.
Collectively, these enhancements deliver the most efficient and secure deployment of QSTT to date, and point the way forward to high-accuracy ultra-secure time transfer in space.


\end{abstract}
\begin{figure*}[htp] 
    \centering
    \includegraphics[width=1\linewidth]{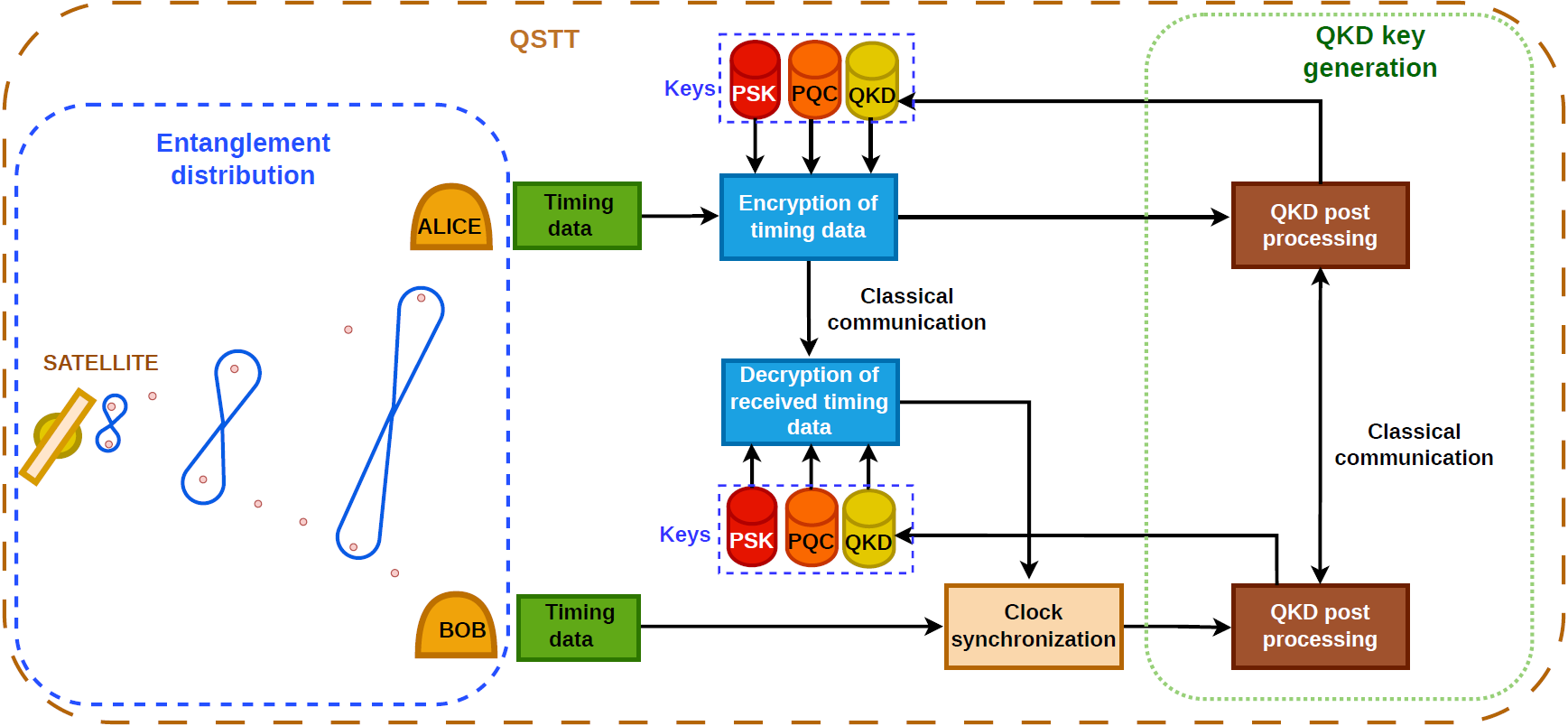}
    \vspace{4pt} 
    \caption{The QSTT system. A satellite with an onboard entangled photon source distributes entangled photons to Alice and Bob's ground stations. Timing data is generated locally by both parties from the photon's detection. Alice encrypts the timing data using the method described in Section~\ref{QSTT} and transmits it to Bob via a classical communication channel. Upon receiving Alice's encrypted timing data, Bob decrypts it, performs clock synchronization using his and Alice's timing information, and corrects his time tag using the estimated clock offset. Upon successful synchronization, they perform QKD post-processing, including key sifting, error estimation, error correction, and privacy amplification, resulting in the QKD keys that are sent to the key store (the cylinder marked as QKD). The cylinders marked PSK and PQC represent the pre-shared key and post-quantum cryptography key stores, respectively, which are used to encrypt the time tag data as discussed in Section~\ref{QSTT}.
    }
    \label{fig:functionality}
\end{figure*}
\section{Introduction}\label{sec:introduction}
Precise time-synchronized remote clocks are essential for a wide range of modern applications, including positioning~\cite{surof2026precise}, telecommunications~\cite{pini2021satellite},
quantum communication~\cite{Kimble2008, doi:10.1126/science.aam9288}, distributed quantum computing~\cite{PhysRevA.59.4249}, and distributed quantum sensing~\cite{Guo2020, Zhang_2021}.
Time synchronization in the classical-only domain is obtained via protocols such as the Global Positioning System~(GPS)~\cite{1537364}, the White Rabbit Protocol~\cite{6070148}, and the Network Time Protocol~\cite{103043}.  However, limited precision and vulnerabilities to attacks such as jamming, time-delays, spoofing, and cyber attacks~\cite{alghamdi2020cyber,alghamdi2021precision} render these protocols fundamentally insecure and performance-limited.

Various quantum clock synchronization protocols and quantum time transfer protocols have been proposed and demonstrated~\cite{PhysRevLett.85.2010, Giovannetti2001, 10.1063/1.5086493, 10.1063/1.5121489, PhysRevA.100.023849,  Quan:22, lafler2023quantumtimetransferpractical, PhysRevApplied.22.024012} to address the limitations and vulnerabilities of classical clock synchronization. In most quantum clock synchronization and quantum time transfer protocols, energy-time entangled photons are used to achieve femtosecond-level precision in estimating the clock offset, leveraging the tight temporal correlation between the birth time of entangled photon pairs~\cite{quan2018experimental}. Entangled photon pairs also ensure security by rendering eavesdropping detectable. For example, an intercept-and-resend attack breaks entanglement correlations and can be detected through Bell-inequality testing or quantum state tomography~\cite{lamas2018secure, alqedra2025entanglementverifiedtimedistributionmetropolitan}, and two-way transmission of quantum signals can mitigate time-delayed attacks~\cite{10.1063/5.0191453, PhysRevA.100.023849}.   
However, security of the photon pairs (the signal)  encoding the timing is insufficient; the security of other protocol data sent over the classical channel  (i.e., the time tags) must be secured otherwise an attacker can forge such data, enabling several attacks~\cite{motero2021attacking, breitinger2025sok, mastromauro2025survey}. Encryption and authentication of protocol data is required to provide complete security. 

A Quantum-Secure Time Transfer Protocol (QSTT) can, in principle, provide the complete security desired~\cite{Dai2020}. QSTT uses quantum signaling for time transfer and secure key generation - the QKD-generated keys are being used to encrypt the time-tag data, with intercept-and-resend attacks addressed via Quantum Bit Error Rate~(QBER) monitoring~\cite{Dai2020, Villas, picciariello2024quantum}. However, we caution that one caveat is that in practice encryption of all timing data using a QKD One-Time Pad (OTP) cannot normally be done since usually the time-tag creation rate exceeds the QKD key generation rate. In~\cite{Dai2020}, QKD-derived keys are used to seed the 128-bit Advanced Encryption Standard~(AES), subsequently used for encryption of time-tag information, thereby bypassing rate limitations. However, AES provides only computational security, not information-theoretic security.

Entanglement-based QKD plays a pivotal role in the realization of entanglement-based QSTT systems. Significant progress has been made in both fiber-based~\cite{liu2024high, zhuang2025ultrabright} and free-space channels towards entanglement-based QKD. Experiments have already demonstrated satellite-based entanglement QKD~\cite{yin2017satellite, Yin2020}. However, performance is limited by the high Earth-satellite channel loss and the low efficiency of entangled-photon-pair generation, which constrains the finite QKD key rate.  Current demonstrations of satellite-based entanglement QKD using two-downlink channels provide a maximum finite QKD key rate of $0.12$~bits~$\text{s}^{-1}$~\cite{Yin2020}.
Clearly, a higher finite QKD key rate would be achieved using a high-generation-rate entangled-photon-pair source. Such a source would also help to achieve higher precision  synchronization, thereby reducing QBER and further increasing the key rate~\cite{huang2025clock,rani2025obfuscatedquantumpostquantumcryptography}.

Here, we implement within our QSTT system a high-generation-rate Entangled Photon Source~(EPS) derived from a \mbox{type-0} Periodically Poled Potassium Titanyl Phosphate~\mbox{(PPKTP)} crystal. We utilize this source's  orders-of-magnitude higher photon-pair generation rate over a \mbox{type-II} PPKTP  crystal ~\cite{steinlechner2014efficient,terashima2018quantum}. As we show, this improvement results in an approximately 30 times increase in  finite key rates and asymptotic key rates compared to the state-of-the-art satellite-based entanglement QKD~\cite{Yin2020}, with the same increase found relative to our previous work that utilized a commercial low-rate entanglement source~\cite{adhikari2025new}.
Also, leveraging the new EPS's advantages, we implement a practical clock synchronization protocol that uses the tight temporal correlation of the entangled photon pairs to achieve sub-nanosecond clock precision, without any dependence on external timing infrastructure. In addition, by using the same quantum resources for QKD and synchronization, we significantly reduce synchronization-related hardware overhead, thereby advancing the feasibility of our system. Furthermore, by integrating post-quantum solutions into our QKD system, we demonstrate a novel entanglement-based QSTT system that provides the most secure QSTT platform currently available. In general, these advances enable our QSTT system to support high-precision global networks and establish a new benchmark for secure satellite-based QSTT.  

The remainder of this paper is as follows: Section~\ref{sec: System model} describes the system model in which an EPS, a QSTT system, and a QKD protocol are discussed; 
Section~\ref{results} presents the results and Section~\ref{Sec:conclusion} concludes the paper.
 
\section{System Model}\label{sec: System model}

\begin{figure*}[htp] 
    \centering
    \includegraphics[width=1\linewidth]{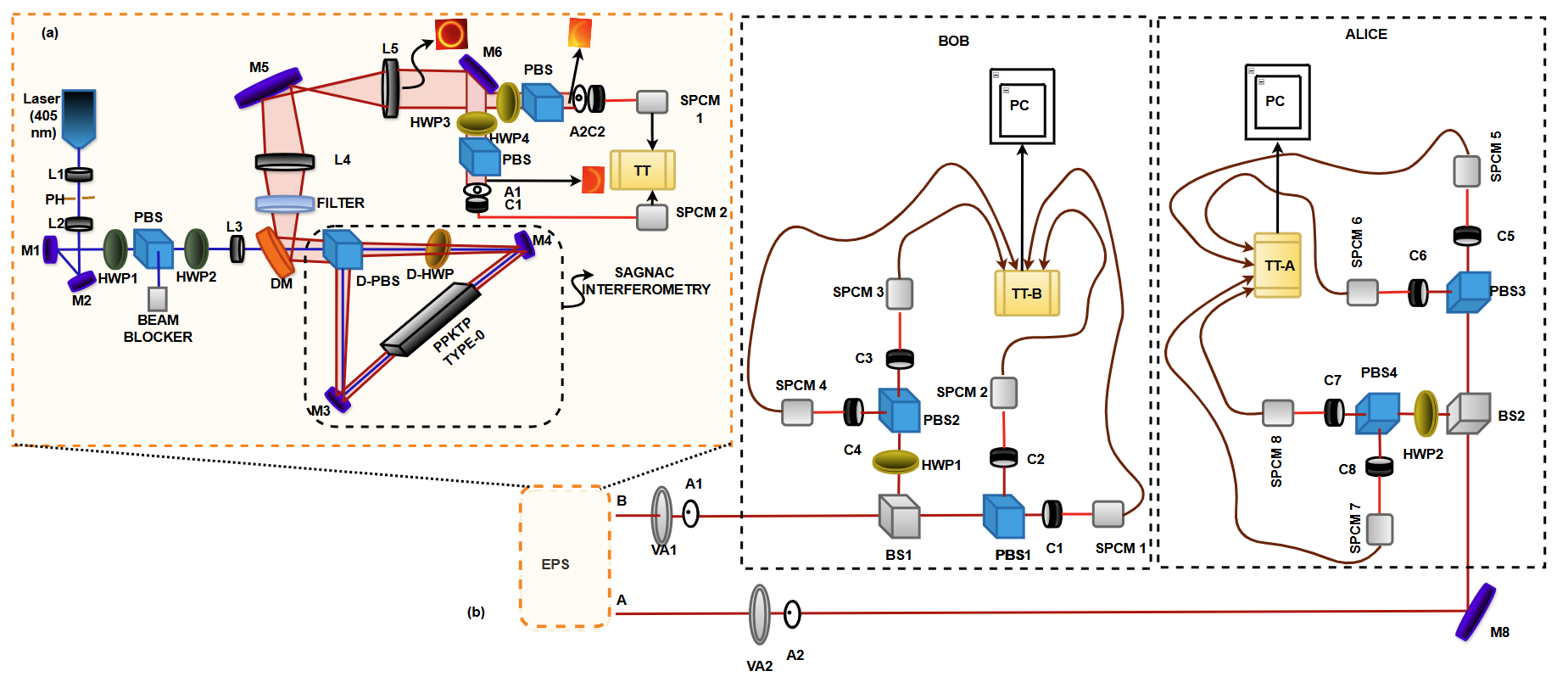}
    \vspace{4pt} 
    \caption{(a) Schematic of the experimental setup for degenerate, non-collinear, polarization-entangled photon-pair generation via the spontaneous parametric downconversion process in a \mbox{type-0} PPKTP crystal using a Sagnac interferometer. The pump and the generated entangled photons are represented by blue and red rays, respectively. PH: Pinhole; HWP: Half-wave plate; PBS: Polarizing beam splitter; DM: Dichroic mirror; SPCM: Single-photon counting module; TT: time-tagger; D-HWP: Dual-HWP; D-PBS: Dual-PBS; M: Mirror; L: Lens; A: Aperture; C: Fiber coupler. 
         (b) Schematic of the QSTT experimental setup. Here, the EPS is used to distribute entangled photons to Alice and Bob's ground stations, each equipped with a measurement setup for the QKD. Points A and B, immediately after the EPS (the dotted rectangular box), indicate the signal and idler paths obtained after removing the visibility-measurement setup shown in Fig.~\ref{fig:full_experimental_setup}(a). Earth-satellite loss emulation is performed using the variable attenuators (VA1 and VA2).
         }
    \label{fig:full_experimental_setup}
\end{figure*}

A high-level overview of our QSTT system is shown
in Fig.~\ref{fig:functionality}. 
We next detail its different features.

\subsection{Entangled photon source} \label{EPS}
For satellite-based quantum communications, the EPS architecture must meet strict size, weight, and power constraints while maintaining performance under temperature variations, mechanical vibrations, and radiations~\cite{anwar2022development, mccarthy2025compact}. Therefore, selecting an appropriate source architecture is critical to the overall performance and reliability of satellite quantum communication systems. Although many interferometric configurations have been designed, such as the Michelson~\cite{katamadze2025broadband}, Mach-Zehnder~\cite{mccarthy2025stable}, Sagnac~\cite{jabir2017robust}, and linear-displacement interferometers~\cite{lohrmann2020broadband}, only a few are suitable for satellite deployment. Sagnac-based EPSs are phase-stable and generate high-quality entangled photon pairs~\cite{anwar2022development}, but they have a large physical footprint. Notwithstanding this, the Micius satellite deployed a \mbox{type-II} PPKTP Sagnac source that enabled quantum communications~\cite{yin2017satellite1, cao2019satellite}. A major challenge in satellite quantum communication is high channel loss~\cite{yin2012quantum, yin2017satellite, Yin2020,lu2022micius}. 
To address this challenge, a \mbox{type-0} PPKTP crystal should be considered over a \mbox{type-II} PPKTP crystal for EPS development, as the \mbox{type-0} nonlinear coefficient in KTP is about $4.7$ times larger than the \mbox{type-II} coefficient, thereby resulting in more than a $22$-fold increase in the photon-pair generation rate~\cite{terashima2018quantum}, defined as the number of photon pairs generated from the crystal per second. An additional enhancement factor of up to four can be achieved in the photon-pair generation rate owing to the large bandwidth of \mbox{type-0} PPKTP spontaneous parametric downconversion~(SPDC)~\cite{steinlechner2014efficient,terashima2018quantum}. 
This results in an approximately two-order-of-magnitude increase in the overall photon-pair generation rate of the \mbox{type-0} crystal compared with that of a \mbox{type-II} crystal of the same length.
The effectiveness of $\mbox{type-0}$ SPDC sources has been demonstrated in prior studies, where a \mbox{type-0} PPKTP crystal in Sagnac interferometry achieved a 1~GHz photon-pair generation rate for a Bell test over an emulated Earth-Moon channel~\cite{cao2018bell}. Similarly, a \mbox{type-0} periodically poled lithium niobate source in~\cite{zhuang2025ultrabright} produced a photon-pair generation rate exceeding $10^{10}$~Hz, enabling entanglement-based QKD over fiber distances of up to $404$~km.

We develop an EPS using a \mbox{type-0} PPKTP crystal in a non-collinear geometry with Sagnac interferometry.
Fig.~\ref{fig:full_experimental_setup}(a) shows the experimental schematic of the polarization-EPS. A single-frequency continuous-wave laser (Toptica, $405.5$~nm wavelength) is used to pump the crystal. 
For spatial filtering of the pump laser, a pinhole arrangement consisting of two $5$~cm focal length lenses (L1 and L2) and a pinhole with a size of $50$~\si{\micro\metre} is used. 
A~$1\times2 \times 30$~mm single-grating \mbox{type-0} PPKTP crystal (Raicol) with a grating period of~$3.425$~\si{\micro\meter} is used.  The crystal is housed in a temperature-controlled oven that maintains the crystal temperature at~$22~\si{\degreeCelsius}$ for non-collinear generation of entangled photon pairs at the degenerate wavelength ($811$~nm). To focus the pump at the center of the crystal, a convex lens L3 with a focal length of~$300$~mm is used. The entangled photon pairs are generated using a dual-half-wave plate~($\text{D-HWP}$) and a dual-polarization beam splitter~($\text{D-PBS}$), two high-quality mirrors (M:~$99$$\%$ reflection coefficient), and the crystal at the center of the Sagnac loop.
The SPDC photons from both clockwise and anticlockwise directions of the Sagnac loop interfere at the $\text{D-PBS}$ and generate the Bell state:~$|\Phi^+\rangle=(|\text{HH}\rangle+|\text{VV}\rangle)/\sqrt{2}$, where $\text{H}$ and $\text{V}$ are the horizontal and vertical photon
polarizations, respectively. 
An $810$~nm filter with a $10$~nm bandwidth is used to transmit only the SPDC photons and block any residual pump photons reflected from a dichroic mirror (DM). 
When imaging the SPDC photons using an electron-multiplying charge-coupled device camera (Andor iXon), a ring is generated; we refer to it as the SPDC ring. A mirror M6 splits the SPDC ring into two equal parts, as shown in Fig.~\ref{fig:full_experimental_setup}(a). The diametrically opposite ends of the ring are selected using the apertures (A1 and A2), and the entangled photons passing through the apertures are coupled into multi-mode fibers and detected using single-photon counting modules (Excelitas AQRH-$14$-FC), henceforth referred to as the photon detectors. The two collection paths are denoted as the signal and idler paths. 
To analyze the polarization-correlation visibility, the measurement setup used for visibility only,  consists of a $\text{PBS}$ and an $\text{HWP}$, is placed in each path before the entangled photons are coupled, as shown in Fig.~\ref{fig:full_experimental_setup}(a). Using this setup, we obtain an average polarization-correlation visibility of~$88$~$\%$~($\text{V}=0.88$), corresponding to an estimated Bell parameter of~$\text{S}=2\sqrt{2}\text{V}~\cite{anwar2021entangled}=2.48$, confirming nonclassical correlations. We also measure the bandwidth ($\Delta\lambda$) of the generated entangled photons using a spectrometer (OceanView) and find $\Delta\lambda$ to be $1.5~$nm. Using $\Delta\lambda$, the width of the temporal correlation between the signal and idler photons is estimated as~$1.5$~ps and is denoted by~$\sigma_{int}$.

\begin{figure}[ht]
		\centering		
        \includegraphics[width=1\linewidth]{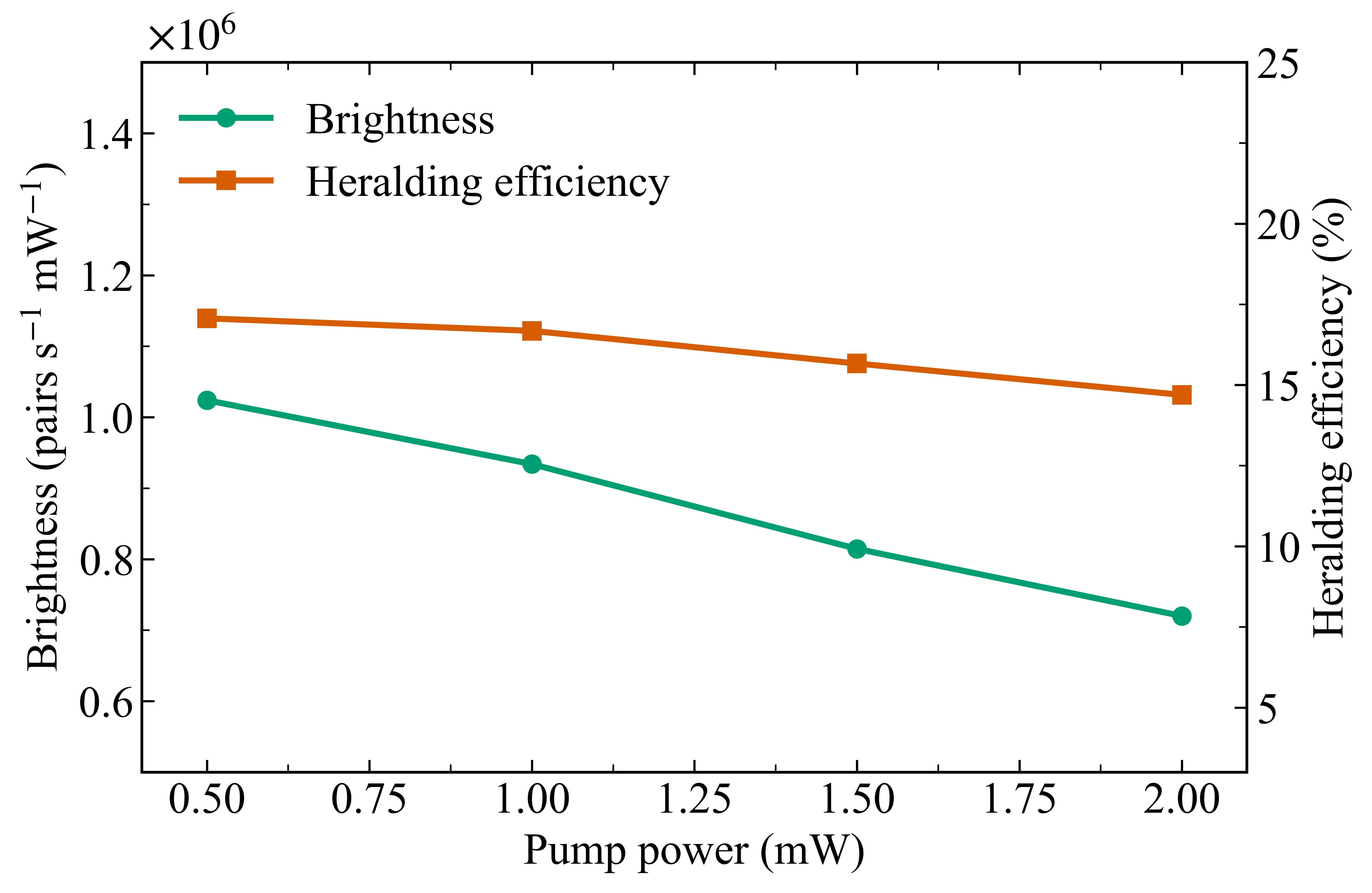}
		\caption{ Measured performance metrics of the EPS. The brightness (on the left y-axis) and the heralding efficiency (on the right y-axis) of the EPS are plotted against pump power.  
        }
        \label{fig:source performance}
\end{figure} 

In Fig.~\ref{fig:source performance}, we present the brightness and heralding efficiency of the EPS as functions of pump power. During these measurements, the entangled photons are collected directly into multimode fibers in both the signal and idler paths, without the visibility measurement setup in place. We define brightness as the coincidence count rate per unit pump power (pairs$~\text{s}^{-1}$$\text{mW}^{-1}$). At an input pump power of $0.5$~mW and using a coincidence window of $1$~ns, the EPS achieves a brightness exceeding $10^{6}$~pairs~$\text{s}^{-1}$$\text{mW}^{-1}$. 
Moreover, we calculate the heralded efficiencies of the EPS, as shown in Fig.~\ref{fig:source performance}, as the average of the signal and idler heralding efficiencies. Heralded efficiency is the measured probability of detecting one photon of a pair when the other is detected. The signal and idler heralding efficiencies are given by: $\zeta^{(1|2)}=\textit{R}_c/\textit{S}_2$ and $\zeta^{(2|1)}=\textit{R}_c/\textit{S}_1$, respectively,
where \(\textit{R}_c\) is the coincidence count rate and \(\textit{S}_1\) and \(\textit{S}_2\) are the count rates in the signal and idler paths, respectively.

\subsection{QSTT System}\label{QSTT}
Our QSTT system enables secure time transfer by combining the QKD and Post-Quantum Cryptography (PQC) architectures, coupled to an obfuscation procedure. The wider motivation and merits of this hybrid approach are detailed elsewhere~\cite{rani2025obfuscatedquantumpostquantumcryptography}. We summarize here its operation, bearing in mind the main contribution of the present work is the development and inclusion of a custom-designed high-generation entangled-pair source, highlighting the many benefits that it brings to almost all aspects of such a system. 

Secret keys generated via QKD and Pre-Shared Keys~(PSKs) are used in the QSTT system to provide information-theoretic security. In the following, we use the terms `QKD-OTP' and `PSK-OTP' to denote one-time padding with the QKD keys and PSKs, respectively. Photon detection time tags are recorded at Alice and Bob's stations over a total measurement time $T_{run}$, in arrays $\boldsymbol{T_A}$ and $\boldsymbol{T_B}$, respectively. The time tags are defined with respect to a global time reference~$O$, and are given by:
\begin{equation}\label{equ1}
t_{\alpha} = t^o + \frac{L_{\alpha}}{c} + \delta t_{\alpha}, \quad \text{where } t_{\alpha} \in \boldsymbol{T_{\alpha}}, \; {\alpha} \in \{A, B\},
\end{equation}
where \({\alpha}\in\{A, B\}\) distinguishes Alice and Bob, $t^o$ is the photon-pair creation time, $L_\alpha$ is the propagation distance,  and \(\delta t_{\alpha}\) denotes the local clock offset relative to $O$. To remove the dependence on $\delta t_A$, Alice's time-tag array is converted into an inter-arrival time (diff-time-tag) array,~$\boldsymbol{\Delta T_A}(j) = \boldsymbol{T_A}({j+1}) - \boldsymbol{T_A}(j),~ j=1,\ldots,n-1,$ where $n$ is the total number of detection time tags over $T_{run}$.
To ensure the confidentiality of the timing data, Alice encrypts the data as follows.
To efficiently use the QKD keys, she applies QKD-OTP selectively. 
(i)~She encrypts her first time tag $ \boldsymbol{T_A}(1)$ using QKD-OTP encryption (ciphertext denoted by $ \boldsymbol{T_A}(1)^*$), since $\boldsymbol{T_A}(1)$ is essential for reconstructing $\boldsymbol{T_A}$ from $\boldsymbol{\Delta T_A}$. 
(ii)~She divides $\boldsymbol{\Delta T_A}$ into $2^h$ partitions, where $h$ denotes the partition depth, and then shuffles the partitions via permutation $\mathcal{P}$, obfuscating the temporal sequence of $\boldsymbol{\Delta T_A}$. The permutation $\mathcal{P}$ is then QKD-OTP encrypted. 
(iii)~She also QKD-OTP encrypts the diff-time tags corresponding to randomly selected indices, $\mathcal{Q}$, from $\boldsymbol{\Delta T_A}$, where the size of $\mathcal{Q}$ is limited by the available QKD keys. 
Since QKD-OTP encryption of the entire diff-time-tag array would consume all of the QKD keys
only diff-time tags\footnote{The maximum number of diff-time tags that can be QKD-OTP encrypted is given by $|Q| \le \left\lfloor \frac{K_{total} - h \cdot 2^h - k_2 - k_{\text{AES}} - k_{1}}{b} \right\rfloor,$ where \( \lfloor \cdot \rfloor \) is the floor operator, $K_{total}$ is the total number of bits available from previous QKD sessions, \( h \cdot 2^h \) is the number of bits consumed in encrypting  $\mathcal{P}$, \( b \) is the number of bits used to represent each diff-time tag, \( k_2 \) is the number of bits consumed for MAC authentication, \( k_{\text{AES}} \) is the number of bits consumed for AES seeding, and \( k_{1} \) is the number of bits consumed to encrypt the timestamp $ \boldsymbol{T_A}(1)$.} indexed by $\mathcal{Q}$ are QKD-OTP encrypted. The encrypted diff-time tags are recorded in an array denoted by $\boldsymbol{\Delta T_A^{\text{OTP}}}$.

While the remaining diff-time tags not QKD-OTP-encrypted, denoted by $\boldsymbol{\Delta T_A^{\text{R}}}$, are protected using an obfuscated encryption-decryption process defined as multiple cascaded encryption (and its corresponding decryption) operations, based on an instruction sequence (IS), applied to data. The IS contains the encryption sequence information. 
The available encryption algorithms include AES-256 with QKD keys and Ascon with keys established via PQC key encapsulation protocols. 
An IS is defined as $ (\xi_1, \xi_2, \dots, \xi_m), \text{where each }
\xi_i \in \{ \text{Ascon}, \text{AES}\}   \text{~and } i = (1, 2, \dots, m)$, where $m$ is the number of times an encryption-decryption protocol is used. For example, an IS may be represented as (Ascon, AES, Ascon). Alice selects an IS from a set of $m'$ instructions to encrypt $\boldsymbol{\Delta T_A^{\text{R}}}$, where $m'$ is the number of ISs set by the system (see Section \ref{results}  for the specific sequences we have used). The selected IS is PSK-OTP encrypted to obtain the ciphertext. Hence, the post-quantum solutions add a pragmatic security layer to the timing data. The encrypted remaining diff-time tag array are represented by $\boldsymbol{\Delta T_A^{\text{IS}}}=\xi_m( \dots \xi_2( \xi_1(\boldsymbol{\Delta T_A ^{\text{R}}})))$. We define $\boldsymbol{T_A}^{*}$ as the final encrypted timing data, expressed as: $~\boldsymbol{T_A^*} = \left(\boldsymbol{T_A(1)^*} , \boldsymbol{\Delta T_A^{\text{OTP}}} , \boldsymbol{\Delta T_A^{\text{IS}}}  \right)$.   

To guarantee the integrity and authenticity of the timing data, an information-theoretic Wegman-Carter message authentication code is applied on $\boldsymbol{T_A^*}$, which consumes some of the QKD keys. Upon successful authentication and decryption, Bob reconstructs $\boldsymbol{T_A}$. Note that "QKD-OTP" is replaced with "PSK-OTP" (except for diff-time tags) when the QKD keys are unavailable. If the PSKs are exhausted, the QKD keys can be used as substitutes. If both are exhausted, the system reverts to using the PQC keys for all functions. Each PSK and QKD bit is used only once.
Hence, the QSTT system provides a secure way to distribute timing information, utilizing QKD resources while protecting all timing data through a combination of information-theoretic and post-quantum security mechanisms. 
 
\subsection{QKD}\label{QKD_protocol}
We implement an entangled version of QKD, specifically BBM92~\cite{bennett1992quantum}, as shown in Fig.~\ref{fig:full_experimental_setup}(b). Entangled photons are distributed to two distant users, Alice and Bob, via two independent free-space quantum channels. In the polarization degree of freedom, the state is~$|\Phi^+\rangle=\dfrac{1}{\sqrt{2}}(|\text{H}\rangle_{\text{S}} |\text{H}\rangle_{\text{I}} + |\text{V}\rangle_{\text{S}} |\text{V}\rangle_{\text{I}})$. Here, subscripts $\text{S}$ and $\text{I}$ denote the signal and idler, respectively. 
After receiving the photons, Alice and Bob measure their polarization using the linear ($\text{H/V}$) and diagonal ($\text{D/A}$) bases. The time-tag arrays are continuously recorded in Alice's and Bob's computers, where post-processing of the QKD protocol is performed. Post-processing includes the transfer of classical data, key sifting, error estimation, error correction, and privacy amplification, resulting in the generation of the QKD keys. A session is completed after privacy amplification, taking a total time $T_{run}$.

In~\cite{Yin2020}, the finite key analysis is performed using the uncertainty approach for smooth entropies~\cite{tomamichel2012tight} with Serfling's bound~\cite{serfling1974probability} and also incorporates deviations in detector efficiencies~(\cite{gottesman2004security}). The finite key length expression used in~\cite{Yin2020} is given by
\begin{align}
\label{eq: Lz}
    L_Z = n_Z-n_ZH\Bigg[\frac{E_X+\sqrt{\frac{(n_Z+1)\log(1/\epsilon_{sec})}{2n_X(n_X+n_Z)}}}{1-\Delta}\Bigg]\\\nonumber-f_en_ZH(E_Z)-n_Z\Delta-\log\frac{2}{\epsilon_{cor}\epsilon_{sec}^2},
\end{align}  
where $n_Z$ and $n_X$ are the number of sifted bits in the $Z$ and $X$ bases, respectively; $E_Z$ and $E_X$ represent the QBER in the $Z$ and $X$ bases, respectively; $f_e$ represents the reconciliation efficiency, and $\Delta$ represents the upper bound on the fractional reduction due to deviations in the photon detector inefficiency. $\epsilon_{sec}$ and $\epsilon_{cor}$ represent the security parameters for secrecy and correctness of the final secret key, respectively. In~\cite{Yin2020}, the $X$ and $Z$ bases measurement blocks are post-processed separately using QBER estimates from their respective test blocks. The finite key lengths $L_Z$ and $L_X$ (similar to Eq.~\ref{eq: Lz}) are then computed, yielding the total finite key length of~$L_Z+L_X$.

In contrast, our implementation provides a single QBER value. Therefore, to fairly compare our finite key rate with~\cite{Yin2020}, we derive a total finite key length $L$ by replacing $n_Z \to \hat{n}$, $E_Z (E_X) \to E (E)$, and $n_X \to k$, where $\hat{n}$ is the number of the key distillation bits, $E$ represents the average QBER, and $k$ represents the number of the test bits.
Hence, the reduced finite key length expression for our work becomes,
\begin{align}
\label{eq: L}
    L = \hat{n}-\hat{n}H\Bigg[\frac{E+\sqrt{\frac{(\hat{n}+1)\log(1/\epsilon_{sec})}{2k(\hat{n}+k)}}}{1-0.0287}\Bigg]\\\nonumber-f_e\hat{n}H(E)-0.971\hat{n}-\log\frac{2}{\epsilon_{cor}\epsilon_{sec}^2},
\end{align}
where we have assumed the deviation in the photon detector efficiencies to be $1.47\%$ (value of $\delta$ from~\cite{Yin2020}).
In addition to using Eq.~\ref{eq: L} for the finite key analysis in our work, we also compute the finite key rate using Clopper-Pearson bounds. We note that in~\cite{Yin2020}, Serfling's bound is used to quantify the bound on the probability of the event that the QBER in the total block (i.e., the test block and the key distillation block combined) lies outside the interval estimated from the test block. In contrast, we use the Clopper-Pearson bounds for the probability that the QBER in the key distillation block is outside the estimated interval obtained from the test block. Further details of the Clopper-Pearson bound implementation can be found in our previous paper~\cite{gupta2026combined}.

\begin{figure*}[htp]
    \centering

    \begin{subfigure}{0.49\linewidth}
        \centering
        \includegraphics[width=\linewidth]{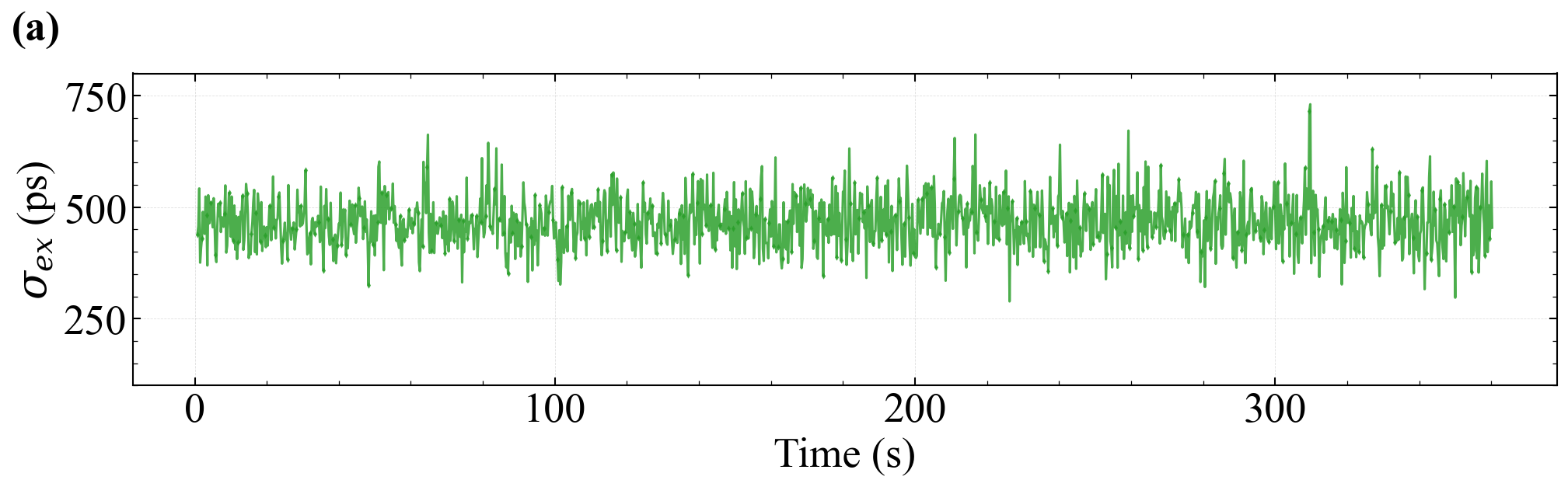}
    \end{subfigure}
    \hfill
    \begin{subfigure}{0.49\linewidth}
        \centering
        \includegraphics[width=\linewidth]{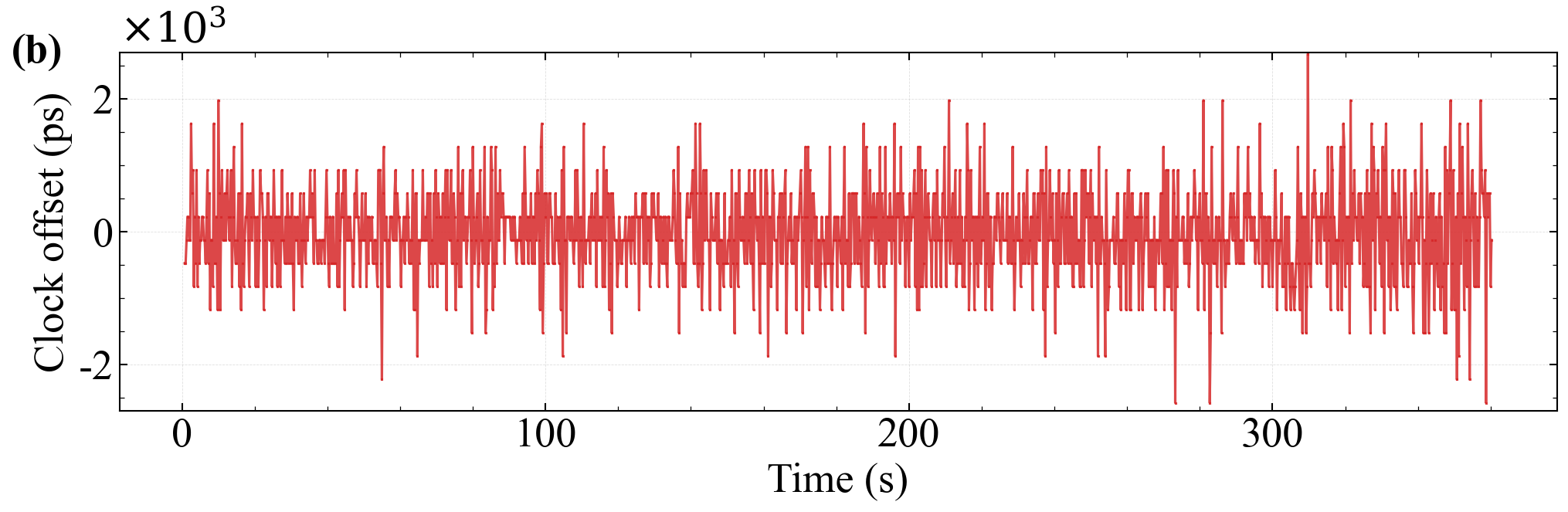}
    \end{subfigure}

    \vspace{6pt}

    \begin{subfigure}{0.49\linewidth}
        \centering
        \includegraphics[width=\linewidth]{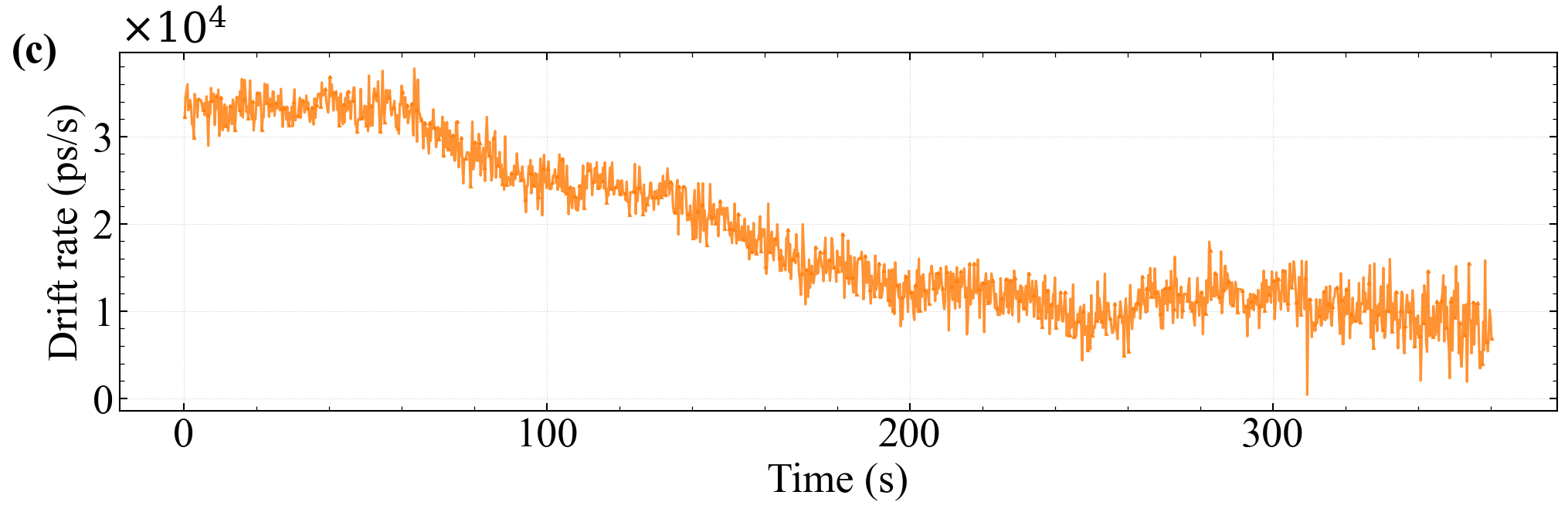}
    \end{subfigure}
    \hfill
    \begin{subfigure}{0.49\linewidth}
        \centering
        \includegraphics[width=\linewidth]{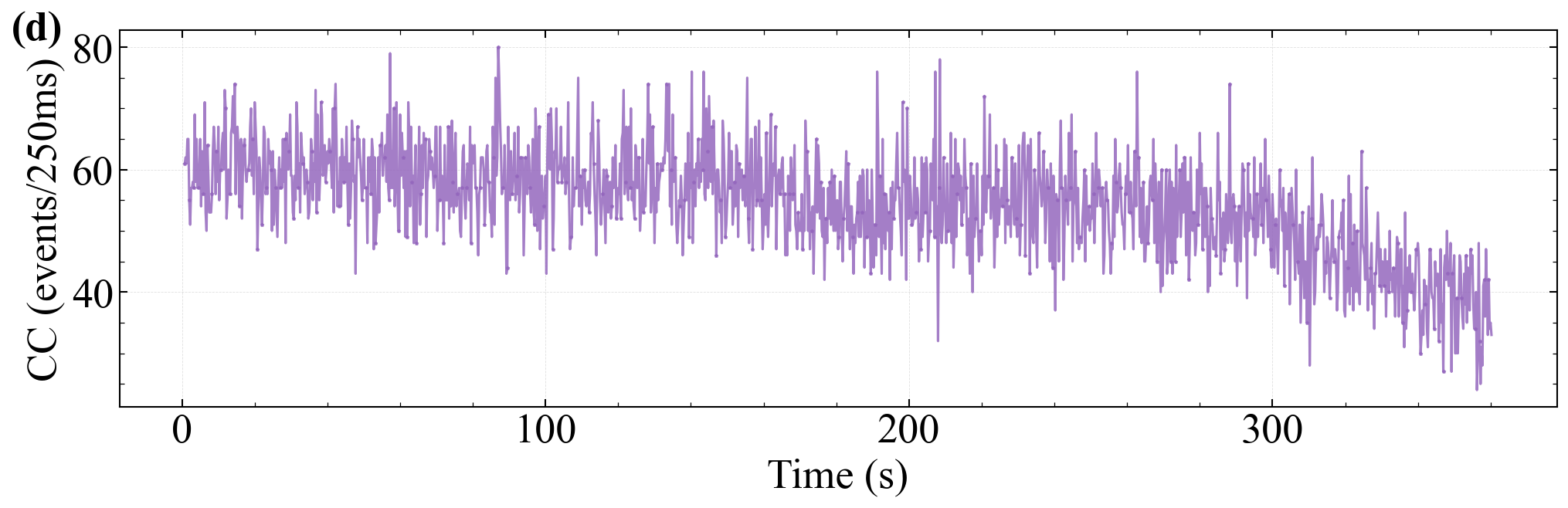}
    \end{subfigure}
     \caption{Clock synchronization protocol execution for the QKD system over 6~minutes of continuous operation at the satellite channel loss. 
    \textbf{(a)}  Experimental system timing jitter $\sigma_{ex}$.
    \textbf{(b)}  Clock offset.
    \textbf{(c)}  Drift rate.
    \textbf{(d)} Coincidence counts (CC) (events/250~ms)), with a 400~ps coincidence window between Alice and Bob.}

    \label{fig:alignment_stability}
\end{figure*}

\section{Experimental Demonstration and Results} \label{results}
Fig.~\ref{fig:full_experimental_setup} shows the experimental setup of our QSTT system. 
The experimental demonstration is performed using the EPS operated at a pump power of $14$~mW over a free-space channel length of 1.2~m, with an overall loss of $63$~dB, hereafter referred to as the satellite channel loss. This loss comprises $32.5$~dB from Alice's channel and $30.5$~dB from Bob's channel. Alice's loss comprises $8$~dB from the coupling and detection losses and $24.5$~dB from A2, while Bob's loss comprises $8$~dB from the coupling and detection losses and $22.5$~dB introduced by A1. The channel loss in our system is verified experimentally to be equivalent to that observed in the Micius satellite over two channels~\cite{Yin2020}. This is achieved by determining the ratio $R_c/G$, where $G$ is the photon-pair generation rate from the source, and $R_c$ is
the coincidence count rate at Alice and Bob's photon detectors. The value of $G$ is determined via $G=S_1S_2/\textit{R}_c$~\cite{tanzilli2001highly}. Using the measured average values of  $S_1=2.5\times10^5~\text{counts s}^{-1}$, $S_2=3.8\times10^5~\text{counts s}^{-1}$, and $\textit{R}_c=216~\text{counts s}^{-1}$, we find ~$G\approx 4.4\times10^8$~pairs~$\text{s}^{-1}$, and a combined average channel loss of $63.1$~dB at a pump power of $14$~mW. Alice and Bob's count rates, as well as the coincidence count rate, exhibited fluctuations throughout the experiment,
with a standard deviation of order $0.75$~dB. 

The QKD-measurement setup employs the photon detectors with a dark count rate of~$100$~s$^{-1}$. The photon detector outputs are connected to a time-tagger (Ultra, Swabian Instruments) with a timing resolution of~42~ps, which timestamps and records photon arrival times for post-processing. The time-tagger uses an internal clock that exhibits significant drift of~$20-70$~ns~$\text{s}^{-1}$. Each QKD session is of~T$_{run}$~=~$3$~minutes.

 In QSTT operational mode, we encrypt Alice's time-tag information using the method described in Section~\ref{QSTT}, with the QSTT parameters: $2^h =~64$, $|Q|=~64$, $m=3$, $m'=4$, and the length of the MAC tag~$l$~=~61. The IS is chosen from the set of four instructions:~$00 \rightarrow (\text{Ascon, AES})$, $01 \rightarrow (\text{AES, Ascon})$, $10 \rightarrow (\text{AES, Ascon, AES})$, and~$11 \rightarrow (\text{Ascon, AES, Ascon})$.
 Alice and Bob use the QKD keys generated between them in previous QKD sessions, as well as the PSKs and the PQC keys.
 It is important to note that the generated QKD-key rate is lower than the time-tag creation rate, rendering QKD insufficient for QKD-OTP encryption of the entire timing data. Therefore, in practice, full information-theoretic security cannot be achieved for the timing data using the self-generated QKD keys.  Despite this limitation, integrating QKD keys across different parts of our QSTT system, alongside PSKs and post-quantum solutions, significantly strengthens the security of the time transfer. We observe that encrypting and authenticating such a large time-tag array introduces a processing delay of a few seconds, which can, in the future, be reduced by high-performance computers.  After a successful QSTT, Bob executes the clock synchronization protocol (discussed in Appendix) to synchronize his clock with Alice's by determining the clock offset and drift through the synchronization algorithms with the parameters provided in Table~\ref{tab:synch_parameters}.
 \begin{figure*}[htp]
    \centering

    \begin{minipage}{0.48\linewidth}
        \centering
        \includegraphics[width=\linewidth]{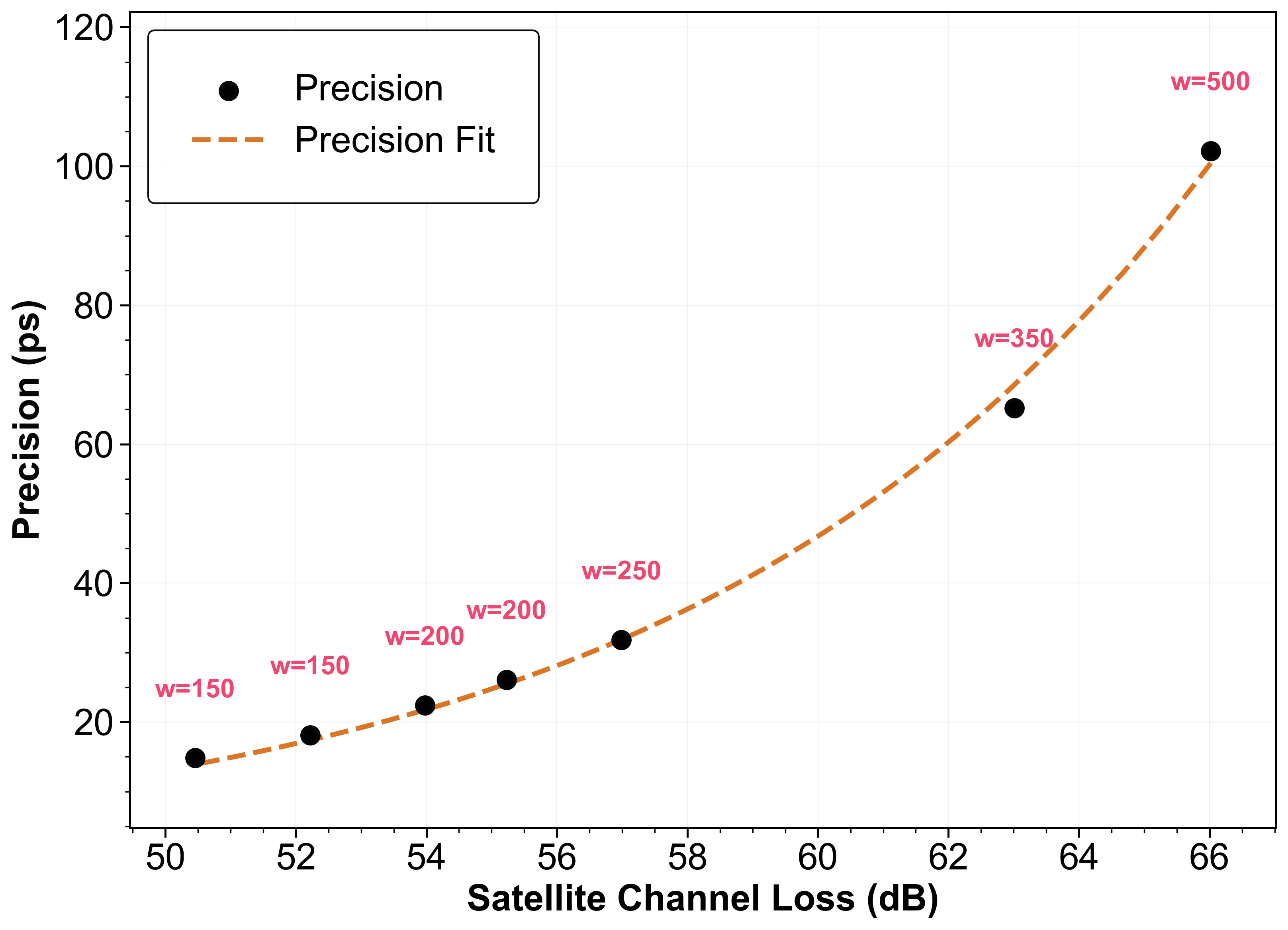}
    \end{minipage}
    \hfill
    \begin{minipage}{0.48\linewidth}
        \centering
        \includegraphics[width=\linewidth]{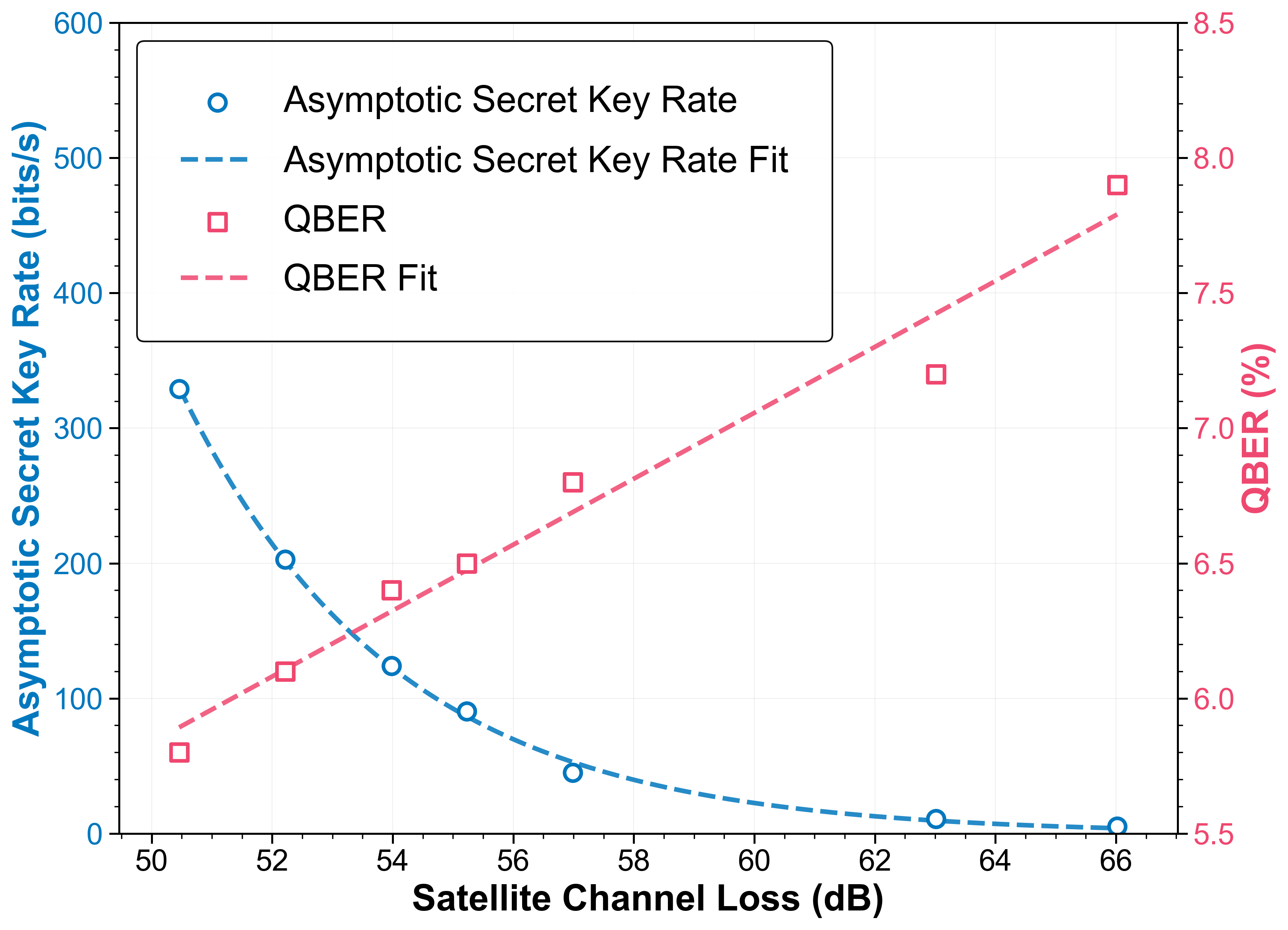}
    \end{minipage}

    \vspace{4pt}

    \caption{Clock synchronization performance of the QKD system under various satellite channel losses.
    (a) Precision versus satellite channel loss analysis plotted with~$w$ values labeled at the top of the scatter points. 
    (b) The asymptotic secret key rate (on the left y axis) and the QBER (on the right y axis) are shown as a function of satellite channel loss. 
    }
    
    \label{fig:combinedQCS_QKD}
\end{figure*}

\begin{table}[h]
\centering
\caption{Synchronization Parameters }
\begin{tabular}{ll}
\toprule
\textbf{Parameter} & \textbf{Value} \\
\midrule
Threshold count rate ($S_{th}$) & 5000~$\text{s}^{-1}$\\
Sampling interval ($\textit{T}_{small}$) & 10~ms \\
Acquisition  period  ($\textit{T}_{acq}$) & 250~ms \\
Coarse scan step size (\( \delta \)) & 100~ns \\
Coarse scan range (\( R \)) & \(\pm10\)~ms \\
Fine alignment bin width (\( w \)) & 350~ps \\
Number of partitions (\( P \)) & 5 \\
QKD coincidence window ($\text{W}$) & 400~ps \\

\bottomrule
\end{tabular}
\label{tab:synch_parameters}
\end{table}

\begin{table*}[t]
\centering
\caption{\textcolor{black}{Comparison of our Quantum Secure Time Transfer (QSTT) system with \cite{Yin2020} under the satellite channel loss of $63$~dB.
}}
\label{tab:qstt_comparison}
\small
\renewcommand{\arraystretch}{2.0}
\begin{tabularx}{\textwidth}{|l|c|X|c|c|c|c|c|c|c|}
\hline
\textbf{Work} &
\textbf{EPS-crystal} &
\textbf{Generation rate} &
\textbf{Av. $\text{CC rate}^{\dagger}$} &
\textbf{$\text{R}_a$} &
\textbf{$\text{R}^s_f$} &
\textbf{$\text{R}^{cp}_f$} &
\textbf{QKD} &
\textbf{Clock synchronization}&
\textbf{QSTT}\\
\hline
~Micius \cite{Yin2020} &
\mbox{Type-II} &
$5.9\times 10^{6}$ at $30$~mW &
2.2&
0.43&
0.12&
$\times$&
\textbf{\checkmark}&
$\times$ &
$\times$ \\
\hline

\hline
\textbf{Our work} &
\mbox{Type-0} &
$4.4\times 10^{8}$ at $14$~mW&
216&
13.13&
3.29&
4.93&
\textbf{\checkmark} &
\textbf{\checkmark} &
\textbf{\checkmark} \\
\hline
\end{tabularx}\vspace{0.1ex}
\raggedright
{\footnotesize
\textcolor{black}{$^{\dagger}$Av. CC rate is the average coincidence count rate received at the ground stations. $\text{R}_a $, $\text{R}^s_a $, and $\text{R}^{cp}_f$  represent the asymptotic key rate, the Serfling's bound-based finite key rate, and the Clopper-Pearson bound-based finite key rate, respectively, in bits~$\text{s}^{-1}$ unit. Clock synchronization refers to synchronization using the temporal properties of entangled photons.  
}}
\end{table*}

In Fig.~\ref{fig:alignment_stability}, we present the performance of the synchronization process over $6$ minutes of continuous operation. The experimental system timing jitter, $\sigma_{ex}=467$~ps (see Fig.~\ref{fig:alignment_stability}(a)). Without the synchronization protocol, $\sigma_{ex}=8$~ns. We provide  additional information about this main synchronization result in the following.

In each synchronization round, the clock offset, $\delta t_{BA}$,  is determined by the position of the correlation peak in the cross-correlation histogram produced by the synchronization algorithms (see the appendix). Its values over many rounds are plotted in Fig.~\ref{fig:alignment_stability}(b).  Despite the variation in drift rate, as shown by Fig.~\ref{fig:alignment_stability}(c), the drift rate is continuously tracked and compensated for by the synchronization algorithm, maintaining an accurate determination of the clock offset. The value of $\sigma_{ex}$ is determined by the width of this correlation peak (assuming a Gaussian fit), and for clarity, Fig.~\ref{fig: fine_Alignment} shows one realization which yields one of the $\sigma_{ex}$ values. The precision of the clock-offset is defined by: $\Delta=\sigma_{ex}/\sqrt{\textit{R}_c\textit{T}_{acq}}$~\cite{Ho_2009,quan2020high}, where $\textit{T}_{acq}$ is the acquisition period. We achieve a precision of $\Delta \approx63$~ps.
We know that $\sigma_{ex}$ depends on the intrinsic temporal width of the biphoton ($\sigma_{int}=1.5$~ps), the photon detector timing jitter ($\sigma_{det}=250$~ps), the time-tagger timing jitter ($ \sigma_{tt}=18$~ps), and the clock drift error ($\sigma_{drift}=\rho^r_{drift}\textit{T}_{acq}=250$~ps) originating from the residual clock drift rate $\rho^r_{drift}$. The value of $\sigma_{ex}$ can be given by: $\sqrt{\sigma_{int}^2 + 2\sigma_{det}^2 + 2\sigma_{tt}^2 + \sigma_{drift}^2 }$, which gives $434$~ps, this is consistent with the $\sigma_{ex}=467\pm56$~ps obtained directly with the cross-correlation width, thereby providing consistency in our estimate of the system timing jitter.

For the QKD process, an average of $54$~coincidence counts (CC) are extracted per~$250$~ms, with a coincidence window of~$400$~ps, as shown in Fig.~\ref{fig:alignment_stability}(d). The synchronization algorithms demonstrate the simultaneous real-time processing of coincidence counts to maintain the QKD performance.  
 In Fig.~\ref{fig:combinedQCS_QKD}, we illustrate the critical importance of precise clock synchronization for the QKD process and quantify the optimal precision that can be achieved at various satellite channel losses for our system. 
 We adjust the fine alignment bin width $w$, which is set based on the average detected coincidence counts, to achieve optimal clock synchronization across different satellite channel losses. In Fig.~\ref{fig:combinedQCS_QKD}(a), we achieve higher precision as the satellite channel loss decreases, which is a direct advantage of the increasing coincidence counts and, in turn, improves the performance of the QKD system, as shown in Fig.~\ref{fig:combinedQCS_QKD}(b). As satellite channel loss increases, precision degrades, which in turn increases QBER and lowers the asymptotic secret key rate, as shown in Fig.~\ref{fig:combinedQCS_QKD}(b).  We observe that at a satellite channel loss of~$66$~dB, synchronization algorithms mostly fail to estimate the true clock offset and clock drift rate due to insufficient coincidence counts.

\begin{figure}[ht]
		\centering		
        \includegraphics[width=1\linewidth]{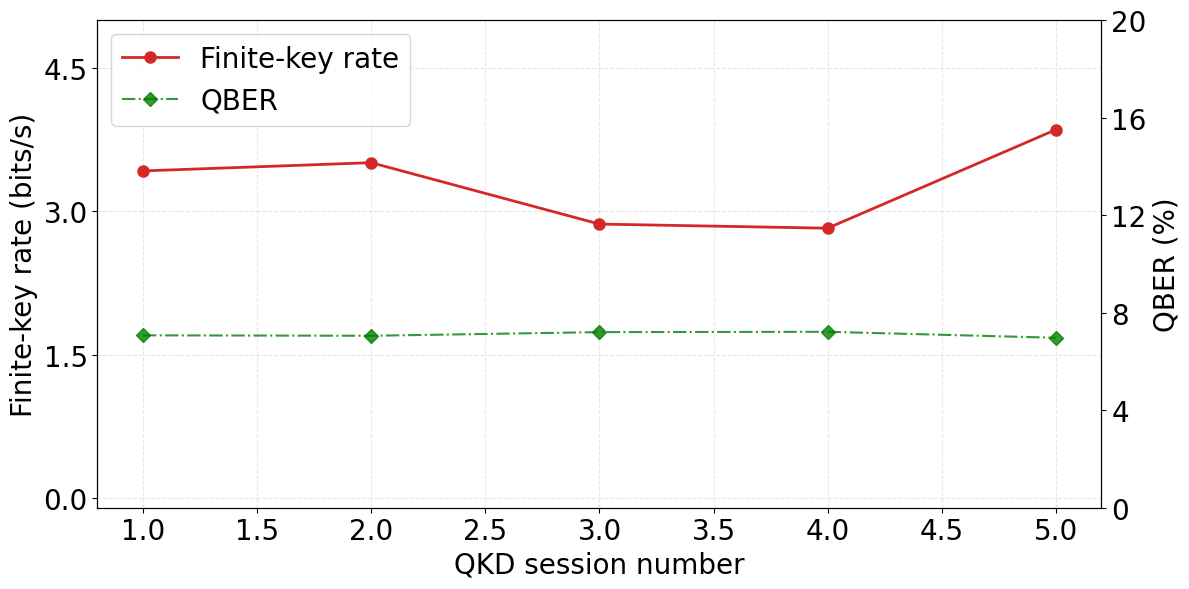}
		\caption{The finite-key rate (using Serfling's bound) and the QBER are measured over $5$~QKD sessions, benchmarking the QSTT performance in the Earth-satellite channel. 
        }
        \label{fig:keyrate_qber_vs_sessions}
\end{figure}

Finally, we demonstrate QSTT (as described in Section~\ref{QSTT}) for each QKD session, under the emulated satellite channel loss. In Fig.~\ref{fig:keyrate_qber_vs_sessions}, we show the finite key rate and the QBER in~$5$ QKD sessions. On average, the QSTT system achieves the asymptotic key rate of~$13.13$~bits~$\text{s}^{-1}$ and the finite key rate (using Serfling's bound) of~$3.29$~bits~$\text{s}^{-1}$ at a QBER value of~$7.1\%$, using parameters $f_e=1.19$, $\hat{n} =k= 10^4$, and $\epsilon_{sec}=\epsilon_{cor}=10^{-10}$. 
The finite key rate and the asymptotic key rate of our system achieve $27$-fold and $30$-fold increases, respectively, compared to \cite{Yin2020}. Table~\ref{tab:qstt_comparison} provides a clear comparison between our system and the Micius-based system~\cite{Yin2020}. Importantly, our system operates at half the pump power used in the Micius source; increasing the pump power to $30$~mW may increase the finite key rate by $60$-fold compared to~\cite{Yin2020}. Furthermore, we compute the Clopper-Pearson bound-based finite key rate and obtain $4.93$~bits s$^{-1}$ for our system. This quantity is not reported in~\cite{Yin2020}, hence a direct comparison is not possible.

\section{Conclusion} \label{Sec:conclusion}
This work demonstrated an entanglement-based QSTT system using a \mbox{type-0} Sagnac-based EPS designed for the Earth-satellite channel. The system integrated our new custom-designed EPS, clock synchronization, QKD, and hybrid post-quantum solutions. We carried out a full QSTT experimental deployment under a total channel loss of~$63$~dB, compared the QKD key rate with that of Micius's entanglement-based QKD~\cite{Yin2020}, and demonstrated approximately a~$30$-fold increase in both the asymptotic and finite key rates. Our QSTT system provided a fully self-synchronized mechanism for clock synchronization, achieving higher precision than the GPS system, thus removing any dependence on external synchronization hardware. Additionally, the results illustrated the important role of clock synchronization performance in achieving higher QKD performance in a real-world, long-distance quantum network. 
These advances place our entanglement-based QSTT system among the most capable experimentally demonstrated QSTT systems to date, particularly in achieving higher QKD key rates, self-synchronized timing, and integrated post-quantum architecture over the Earth-satellite channel.
The work reported here paves the way for scalable implementation of entanglement-based QSTT systems and holds great promise for practical synchronization in Earth-satellite channels.

\section*{\label{sec:Acknowledgments}Acknowledgments}
The authors thank Dr. Dushy Tissainayagam from Northrop Grumman Australia (NGA) for useful discussions. 
This research has been carried out as a project co-funded by NGA and the Defense Trailblazer Program, a collaborative partnership between the University of Adelaide and the University of New South Wales, co-funded by the Australian Government, Department of Education.

\appendix

\section{Clock synchronization protocol}\label{rob3} 

\begin{algorithm}
\caption{Fine Alignment}\label{fine_algo_drift}
\SetAlgoLined
\DontPrintSemicolon
\SetKwComment{Comment}{/* }{ */}

\KwIn{The time tag array of Alice $\hat{t}_a$:  the time tag array of Bob $\hat{t}_b$:
 the coarse offset $\tau_{\text{coarse}}$: the cumulative offset $\tau_{\text{accum}}$:
the coarse search step size $\delta$: the bin width $w$: the cumulative drift rate $\rho^c_{\text{drift}}$: the acquisition  period  $\textit{T}_{acq}$; the number of partitions of the coincidence array \( P \): the QKD coincidence window: $\text{W}$}

\KwOut{Fine offset $\tau_{\text{fine}}$: the filtered coincidence-index arrays $(\hat{I}_a^{\dagger}, \hat{I}_b^{\dagger})$: the experimental system timing jitter $\sigma_{ex}$: $\tau_{\text{accum}}$: $\rho^c_{\text{drift}}$}

\BlankLine

$\hat{t}_b \gets \hat{t}_b + \tau_{\text{coarse}} + \tau_{\text{accum}} + (\hat{t}_b-\hat{t}_b[0])\cdot\rho^c_{\text{drift}}$\;

\BlankLine

$[\mathcal{C}, \hat{I}_a, \hat{I}_b] = \{[(\hat{t}_a^{(i)}, \hat{t}_b^{(j)}), i, j] : |\hat{t}_a^{(i)} - \hat{t}_b^{(j)}| \leq \delta/2\}$\;

$\mathcal{N}\gets |\mathcal{C}|,\quad n' \gets \lfloor \mathcal{N}/P \rfloor$\;

Partition $\mathcal{C}$ into $P$ segments $\{\mathcal{C}_1,\ldots,\mathcal{C}_P\}$\;

$\mathcal{P} \gets \{\}$\;

$\hat{I}_a^{\dagger},\hat{I}_b^{\dagger} \gets \{\},\{\}$\;

\For{$p = 1, 2, \ldots, P$}{

$\Delta \hat{t}_{ba} \gets \hat{t}_b - \hat{t}_a, \quad \forall (\hat{t}_a, \hat{t}_b) \in \mathcal{C}_p$\;
$b \gets \text{ Histogram}(\Delta \hat{t}_{ba},w)$\;
$\tau_p \gets w\cdot(\arg\max_b |\{j : b_j = b\}|)$\; 

$\mathcal{P} \gets \mathcal{P} \cup \{\tau_p\}$\;

 $(\hat{I}_a^{\dagger}, \hat{I}_b^{\dagger}) \gets$ store indices where $|\Delta \hat{t}_{ba} - \tau_p| \leq \text{W}/2$\;

}
\BlankLine
\CommentSty{/* Compute clock drift rate */}

 $\mathcal{T} \gets \{\hat{t}_b[\hat{I}_b[pn'/2]]-\hat{t}_b[\hat{I}_b[n'/2]] : p = 1, \ldots, P\}$\;

$\rho_{\text{drift}} \gets  \mathrm{slope}(\mathcal{T}, \mathcal{P})$

\BlankLine
\CommentSty{/* Compute fine offset, clock offset, update cumulative offset, update drift rate */}

$\tau_{\text{fine}} \gets \text{mean}(\mathcal{P})$\;
$\delta t_{BA} \gets \tau_{\text{coarse}}+\tau_{\text{fine}}$\;

$\rho^c_{\text{drift}} \gets \rho_{\text{drift}} + \rho^c_{\text{drift}}$\;
$\tau_{\text{accum}} \gets \tau_{\text{accum}} +\delta t_{BA}$ + $\textit{T}_{acq}$$\rho^c_{\text{drift}}$\;
$\tau_{\text{coarse}} \gets 0$\;
\BlankLine
\CommentSty{/* Estimate experimental system timing jitter via Gaussian fitting */}

 $\Delta t \gets \{\hat{t}_b^{(k)} - \hat{t}_a^{(l)} : k\in \hat{I}_b, l \in \hat{I}_a \}$\;
 $c \gets \text{ Histogram}(\Delta \hat{t},w)$\;
 $*,\sigma_{ex} \gets \text{GaussianFit}(c)$\;
{\tiny \CommentSty{Note: For the first round,$\tau_{\text{accum}}$ and $\rho^c_{\text{drift}}$ are initialized to zero. }}

\end{algorithm}
The clock synchronization protocol uses timing data to accurately align two clocks by updating any clock offset that may be present. In our QSTT system, the high-generation-rate EPS ensures that a sufficient coincidence count rate is detected despite Earth-satellite channel losses, thereby making clock synchronization reliable. 
From Eq.~(\ref{equ1}), we form:
$\Delta t_{BA} \equiv t_B - t_A = \frac{L_B - L_A}{c} + \delta t_{BA}$, where \( \delta t_{BA} = \delta t_B - \delta t_A \) is the clock offset. Provided $L_A$ and $L_B$ are known\footnote{Although our synchronization protocol is a one-way transfer scheme, two-way schemes could be implemented with similar security at the cost of additional hardware.}, $\delta t_{BA}$ can be determined. In practice, however, the clock offset is not a constant but a time-dependent quantity due to imperfections in the clocks. We refer to the change of clock offset with time as the clock drift, which is characterized by the drift rate given by:~$\rho_{\text{drift}}=\frac{d\delta t_{BA}}{dt}$. Therefore, to maintain clock synchronization, the clock offset and drift rate must be estimated accurately. 

Our clock synchronization protocol comprises two steps that estimate a clock offset and drift rate between Alice and Bob. The first step is a data packet alignment, performed only once at the beginning of the synchronization protocol to estimate a rough offset with millisecond-level accuracy, ensuring that Alice and Bob's time-tag arrays have correlated events. To achieve this, the EPS is initially blocked, and the background count rate is measured. The threshold count rate ($S_{th}$) is then empirically defined as approximately $25\%$ above the background count rate. The threshold acts as a switch that determines whether the synchronization process should start. At Alice and Bob's stations, a counter-function algorithm records the number of photons detected every~$\textit{T}_{small}$~(the sampling interval). When the EPS is unblocked, the detection count rate exceeds $S_{th}$. The stations then continuously record timing data over an acquisition period of~$\textit{T}_{acq}$, generating time tag arrays $\hat{t}_a$ and $\hat{t}_b$ for Alice and Bob, respectively. Bob estimates the rough offset by taking the difference between Alice's first-time tag and Bob's first-time tag.
The alignment step of the data packet is similar to that reported in \cite{pelet2025entanglement}.
\begin{figure}[ht]	
        \includegraphics[width=1\linewidth]{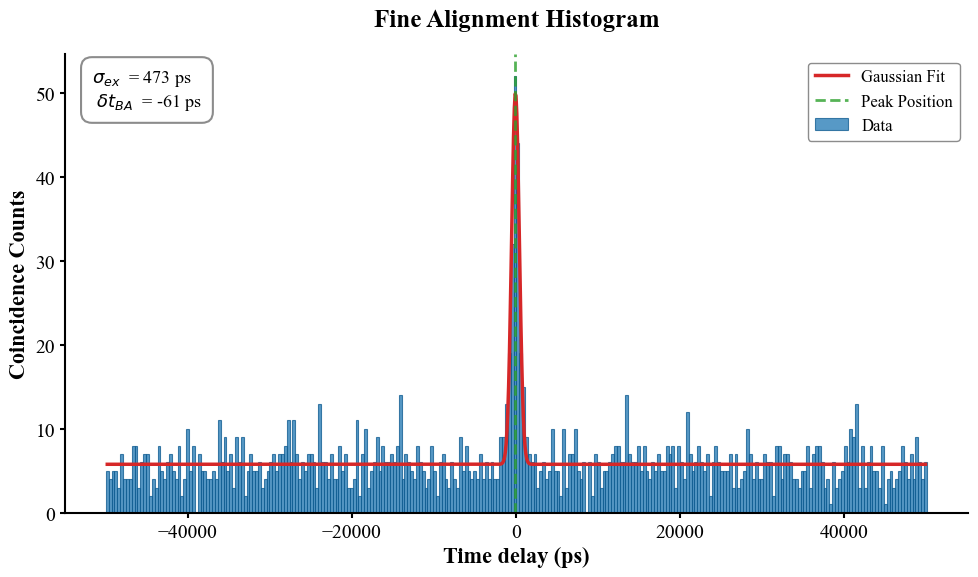}
        \caption{Fine alignment histogram with Gaussian fit after corrections. The red curve shows the Gaussian fit, and the green dashed line shows the peak position.
        Fitted parameters are $\sigma_{ex} = 473$~ps (experimental system timing jitter, i.e., width of the correlation peak) and
        $\delta t_{BA} = -61$~ps (clock offset at the correlation peak), with a bin width $w$ of~$350$~ps.}
        \label{fig: fine_Alignment}
\end{figure}

Following the data packet alignment step, Bob performs the second step, which is the execution of the synchronization algorithms, which comprises the coarse alignment algorithm described in \cite{rani2025obfuscatedquantumpostquantumcryptography} and the fine alignment algorithm (Algorithm~\ref{fine_algo_drift}), on $\hat{t}_a$ and $\hat{t}_b$. The coarse alignment algorithm provides a coarse offset $\tau_{\text{coarse}}$ with an accuracy $\delta$ through a cross-correlation scan over the range $R$ with step size $\delta$. The fine alignment algorithm then precisely estimates the fine offset $\tau_{\text{fine}}$ and drift rate $\rho_{\text{drift}}$. The fine alignment algorithm starts by correcting Bob's time tags based on the coarse offset $\tau_{\text{coarse}}$, cumulative offset $\tau_{\text{accum}}$ (previous accumulated clock offset), and cumulative drift rate $\rho^c_{\text{drift}}$ (previous accumulated drift rate). Coincidence events $\mathcal{C}$ are collected using a coincidence window $\delta$ and partitioned into $P$ segments ($P$ referred henceforth to as the number of partitions of the coincidence array) from which the local offset $\tau_p$ is estimated via correlation peak positions. 
The fine offset $\tau_{\text{fine}}$ is obtained as the mean of the local offsets, while the drift rate is estimated from the slope of a linear fit to the local offsets as a function of time. 

 Finally, the cumulative offset $\tau_{\text{accum}}$ and the cumulative drift rate $\rho^c_{\text{drift}}$ are updated at each synchronization round (a synchronization round is one execution of the synchronization algorithms), and the clock offset $\delta t_{BA}$ is obtained by summing the coarse and fine offset.

In addition to maintaining clock synchronization,  the fine alignment algorithm directly provides the filtered coincidence-index arrays, $\hat{I}_a^{\dagger}~\text{and}~\hat{I}_b^{\dagger}$, from Alice and Bob's time tag arrays within a coincidence window $\text{W}$, referred to as the \mbox{QKD coincidence window}, for the QKD process.
The algorithm is designed for large drifting quantum communication systems to ensure stable synchronization.
Our clock synchronization protocol is easy to implement, computationally efficient, and highly parallelizable. Only the first synchronization round takes approximately $30$~seconds due to the additional execution of the coarse alignment algorithm, whereas subsequent synchronization rounds require only a few milliseconds. In comparison, the state-of-the-art synchronization protocol in~\cite {PhysRevApplied.19.054082} takes approximately $2$~hours to obtain the first clock offset, making it less suitable for real-time clock synchronization over the Earth-satellite channel. This significant reduction in synchronization latency in our clock synchronization protocol enables real-time synchronization within the QKD, offering robustness and flexibility under real-world conditions.

{\small
\bibliographystyle{IEEEtran}
\bibliography{IEEEabrv,main}
} 
\end{document}